\newcommand{\grad}{ {\bf \nabla } }

\newcommand{\EQ}{\begin{equation}}
\newcommand{\EN}{\end{equation}}
\newcommand{\EQA}{\begin{eqnarray}}
\newcommand{\ENA}{\end{eqnarray}}

\newcommand{\bm}{\boldmath}

\newcommand{\JJ}{\mbox{\boldmath $j$} {}}
\newcommand{\uu}{\mbox{\boldmath $u$} {}}

\newcommand{\De}      {\mathrm{D}}

\newcommand{\Div} {{{\bm \nabla}}{\bm \cdot}}

\newcommand{\Laplace} {\nabla^2}

\newcommand{\Strain}{\mbox{\boldmath ${\sf S}$} {}}

\newcommand{\Av}      {\mbox{\boldmath $A$} {}}
\newcommand{\Bv}      {\mbox{\boldmath $B$} {}}

\documentclass{aastex631}
\usepackage{amsmath}
\usepackage{booktabs}
\usepackage{hyperref}
\usepackage{bm}
\usepackage{mathtools, nccmath}
\usepackage[b]{esvect}

\usepackage{color}  
\usepackage{soul}       
\usepackage[normalem]{ulem}
\newcommand{\bcr}{\bf\color{red}} 

\begin{document}

\title{Dependence of spicule properties on magnetic field -- results from Magnetohydrodynamics simulations}
\correspondingauthor{Chatterjee, Piyali}
\email{piyali.chatterjee@iiap.res.in}

\correspondingauthor{Dey, Sahel}
\email{sahel.dey@newcastle.edu.au}

\author[0009-0009-7069-5729]{Kartav Kesri}
\affiliation{Indian Institute of Science Education and Research - Bhopal, Bhopal Bypass Road, Bhauri, Bhopal, 462066, India}
\affiliation{Indian Institute of Astrophysics, Koramangala, Bengaluru, 560034, India}

\author[0000-0002-3369-8471]{Sahel Dey}
\affiliation{School of Information and Physical Sciences, University of Newcastle, University Drive, Callaghan, NSW 2308, Australia}
\affiliation{Indian Institute of Astrophysics, Koramangala, Bengaluru, 560034, India}


\author[0000-0002-0181-2495]{Piyali Chatterjee}
\affiliation{Indian Institute of Astrophysics, Koramangala, Bengaluru, 560034, India}

\author[0000-0003-3439-4127]{Robertus Erdelyi}
\affiliation{Solar Physics and Space Plasma Research Centre (SP2RC), School of Mathematics and Statistics, University of Sheffield, Hicks Building, Hounsfield Road, Sheffield, S3 7RH, UK}
\affiliation{Department of Astronomy, Eötvös Loránd University, 1/A Pázmány Péter sétány, H-1117 Budapest, Hungary}
\affiliation{Gyula Bay Zoltán Solar Observatory (GSO), Hungarian Solar Physics Foundation (HSPF), Petőfi tér 3., Gyula, H-5700, Hungary}

\begin{abstract}

Solar spicules are plasma jets observed in the interface region between the visible solar surface and the corona. At any given time, there is a forest of spicules originating in the chromosphere of the Sun. While various models attempt to elucidate their origin and characteristics, here, we consider the one driven by the magneto-convection undulations. The radiative magneto-hydrodynamical (rMHD) equations are solved using {\sc Pencil Code} with a spatial resolution of 16 km using various magnetic field strengths. The obtained rMHD simulation data are investigated to unveil the various trends in spicular properties as a function of the applied magnetic fields. The important outcome of this study is the finding of a consistent reduction in both the number density and the maximum height reached by spicules as magnetic field strength increases. We also use parabolic fitting on time-distance curves of spicules that are taller than $75^\mathrm{th}$ percentile in the distribution, in order to find a relation between the deceleration of the spicule tip and the magnetic field strength. Our results offer insights into the response of solar spicules to magnetic field strength.

\end{abstract}


\section{Introduction} \label{sec:intro}

Spicules, first discovered by \cite{secchi1877} and later named spicules by \cite{roberts1945preliminary}, are elongated plasma jets (4000-20,000\,km in height), a forest of which is observed all over the chromosphere \citep{zhang2012revision, beckers1972solar, sterling2000solar, tsiropoula2012solar}. Spicules and macrospicules are considered as potential candidates for transporting material and kinetic energy to the corona \citep{woltjer1954photometric,rush1954recent, li2009measurements}. There has been a great interest in spicular jets as a possible source of chromospheric heating and an important component of the chromosphere. If only a small fraction of the mass flux carried by spicules to the corona is able to escape, it may be sufficient to sustain the solar wind \citep{thomas1961physic}. 
So far, models based on these drivers have not been able to quantitatively match both the heights and abundance of the observed solar spicules. There are several mechanisms that are invoked as drivers, for example, rebound shocks \citep{hollweg1982rebound}, Alfv\'en pulses and waves \citep{liu2019evidence,scalisi2021a,oxley2020}, solar global ($p$-mode) oscillations 
 \citep{DePontieu04}, magnetic reconnection \citep{sterling1991numerical,shibata2007chromospheric,samanta2019generation},  magnetic tension release \citep{martinez2017generation} and the Lorentz force \citep{iijima2017three} to name a few, to produce jets whose heights agree better with observations. 
 
Recently, \cite{dey2022polymeric} have argued that solar convection below the photosphere can drive a forest of spicules with a distribution of heights between 5-25\,Mm. Convection in the presence of magnetic fields, also known as magneto-convection \citep{haerendel1992weakly, liu2019evidence} leads to undulations in the photospheric layer of the Sun, which in turn generates slow MHD wave pulses that grow in amplitude because of the sharp density stratification along the vertical direction. This non linear steepening of MHD wave pulses propel the lighter plasma of the chromosphere upwards to form jets. Now, these perturbations cause various structures to form, such as dynamic fibrils, mottles, and spicules \citep{hansteen2006dynamic} or even macrospicules. 
There has been some observational evidence that dynamic fibrils found in and near active regions could be shorter in height than mottles found in the Quiet sun region or spicules that are coronal hole counterparts of the same jets \citep{Shibata1982,DePontieu04}. More recent works like \cite{zhang2012revision} and \cite{Narang2016} analysed Hinode and IRIS data, respectively, and proposed that Coronal Hole jets appear to be faster and longer than those in the Quiet Sun, even though their generation mechanism appears to be the same by the intensity of bright points at their bases. For macrospicules, a statistical study carried out for their properties using SDO/AIA observations detected in Active Regions, coronal holes and Quiet Sun, has shown their modulation as function of the overall magnetic field  \citep[see e.g.]{kiss2017macro}. They explored macrospicular properties (e.g., height, length, average speed, and width) and their dependencies on overall magnetic fields manifested by coronal holes and quiet solar regions, in the first half of Solar Cycle 24. Instead of macospicule types, they found long-term oscillatory behaviour in some properties. 

Here, we perform a statistical study of simulated solar spicules formed due to the mechanism of quasi-periodic driving of the solar convection \citep{dey2022polymeric}, by solving the radiative MHD (rMHD) equations using the {\sc Pencil Code}\footnote{https://pencil-code.org/}. We generate a set of simulations by varying the imposed magnetic field \citep{chatterjee24}\footnote{The MHD setup is publicly available on Zenodo: \url{https://doi.org/10.5281/zenodo.11481216} } in the domain to gain insight on whether the ambient magnetic field modulates the height of these chromospheric jets. 
{\bf For a comprehensive description of the saturation process of the imposed magnetic fields and the distribution of simulated spicules compared to the observed super-granular lanes, readers are encouraged to follow the method section of \cite{dey2022polymeric}.}
This article is divided as follows: under section \ref{S-Methodology} we discuss the methodology we have adopted, under section \ref{S-Results}, we discuss our findings from our analysis and conclude with further discussions in section \ref{S-Discussion}.

\section{Methodology}
    \label{S-Methodology}
    \subsection{Radiative MHD simulations of the solar atmosphere}
    Here, we solve the equations of fully compressible magneto-hydrodynamics (MHD) for a 2-dimensional domain representing the solar subsurface convection, photosphere and the lower solar atmosphere above. The horizontal extent of the domain is 18\,Mm and the total vertical extent including 5\,Mm of convection zone is 37\,Mm. We use a uniform grid size of 16\,km in each direction. While the model has been described extensively in \cite{chatterjee20} and \cite{dey2022polymeric}, we provide the framework used for completeness. Starting from the mass conservation equation, 
    \begin{equation}
\label{eq:continuity}
\frac{\De\ln \rho}{\De t} \,=\,-{\mathbf{\nabla}\cdot}\uu,
\end{equation}
where, $\rho$ is the local plasma density in units of the photospheric density, $\rho_0=2.7\times10^{-7}$ g cm$^{-3}$. 
The Navier-Stokes equation for the magnetized and compressible fluid velocity, $\uu$,  is
\EQ
    \frac{\De\uu}{\De t}\,
   =\,-\,\frac{\grad p}{\rho}   
      + g_z {\hat{z}}                         
      + \frac{\JJ\times\Bv}{\rho}+\mathbf{F}^\mathrm{corr}_L  
      + \rho^{-1}\mathbf{F}_\mathrm{visc}\,,
\label{eq:NS}
\EN
where $\JJ$ is the current density, $\Bv$ is the magnetic field, $\mathbf{F}^\mathrm{corr}_L$ is the semi-relativistic correction to the Lorentz force due to \cite{boris1970nrl} that has been discussed in detail in  \cite{chatterjee20}. The viscous force is modelled as
\begin{align}
\rho^{-1}\mathbf{F}_\mathrm{visc}\, =\, &\,\nu \bigl( \Laplace\uu + \tfrac{1}{3}{ \nabla}\Div\uu
      + 2\Strain {\cdot}{ \nabla}\ln\rho\bigr) \nonumber \\ &\,+\zeta_\mathrm{shock}\bigl[{ \nabla} \left(\Div\uu\right)+\left({ \nabla}\ln\rho+{ \nabla}\ln\zeta_\mathrm{shock}\right)\Div\uu \bigr]\,,
\label{eq:visc}
\end{align}
where, $\nu$ is the kinematic viscosity, and $\Strain$ is the traceless rate-of-strain tensor. We also use an enhanced viscous force at the shock fronts. We use an equation of state that includes ionization calculated assuming {\em{local thermal equilibrium}} (LTE). The pressure, $p$, is given by
\[
p\,=\,\frac{\rho R_g T}{\mu(T)}\,,
\]
where, $R_g=k_B/m_u$ is the ideal gas constant, $T$, is the temperature and $\mu$, is the effective mass given by
\[
\mu\,=\,\frac{4 x_\mathrm{He}+1}{y_\mathrm{H}(T)+x_\mathrm{He}+1}\,.
\]
We consider the number fraction of Helium, $x_\mathrm{He}$, to have a constant value of 0.089, whereas the fraction of ionized Hydrogen, denoted by $y_\mathrm{H}$ is a function of temperature. To calculate $y_\mathrm{H}(T)$ at each time step Saha's ionization approximation is used. For further details, the reader is referred to \cite{chatterjee20}. It has also been shown in \cite{dey2022polymeric}, using the method of Lagrangian tracking, that the dominant term in the Eq.~\ref{eq:NS} that contributes to the upward acceleration of the plasma is the $(-\nabla p/\rho+g_z\hat{z})$ term. The Lorentz force in the same equation seems to play a subsidiary role by channeling the plasma along the field lines.

The induction equation is solved for the magnetic vector potential, $\Av$, using the uncurled induction equation
\begin{equation}
\label{eq:induc}
  \frac{\partial\Av}{\partial t}\,
  = \,\uu\times\Bv - \eta\mu_0\JJ +{\bm \nabla}\varPsi\,.
\end{equation}
In presence of an imposed vertical magnetic field, $B_\mathrm{imp}$, $\Bv=B_\mathrm{imp}\hat{\mathbf{z}}+\Bv'$ and ${\bm \nabla}\times\mathbf{A} = \Bv'$ and $\eta$ denotes molecular magnetic diffusivity. Gauge freedom allows us to set $\varPsi=0$ (Weyl gauge) at all times. 
 
The initial stratification of temperature is obtained by collating the Model S \citep{christensen1996current} for the interior and the atmospheric model by \cite{vernazza1981structure}{\bcr .} The initial density stratification corresponding to temperature is obtained by solving the hydrostatic balance subjected to the ionized ideal gas equation of state with ionization fraction given by the Saha-ionization model. 
 
Finally, we have for the temperature equation, with turbulent diffusion, $\chi_t$,
\begin{align}
   \rho c_V T\frac{\De \ln T}{\De t}
 \,  =\,&\,  -\,(\gamma-1)\rho c_v T \Div \uu+\Div(\mathbf{q}_\mathrm{cond}+\mathbf{q}_\mathrm{rad}) + \Div(\rho T \chi_t {\bm \nabla} \ln T)\nonumber\\
     &\, + \eta\mu_0\JJ^2
      + 2\rho\nu{\sf{S}}_{ij}^2 +{\mathbf \rho\zeta_\mathrm{shock}\left({\bm \nabla}{\bm \cdot}\uu\right)^2}
      -\rho^2\varLambda(T) \,.\label{eq:entropy}
\end{align}
The Spitzer heat conduction flux is denoted ${\mathbf{q}}_\mathrm{cond}$ and described in \cite{chatterjee20},
$\JJ$ presents the current density and is associated with the Ohmic heating, ${\sf{S}}^2_{ij}$ denote the square of the tensorial components of the rate-of-strain tensor, $\Strain$ summed over all indices, $i, j$ and, $c_V$ is the specific heat capacity at constant volume. 
The radiative flux is denoted by $\mathbf{q}_\mathrm{rad}$ and is calculated using the method of long characteristics as described in \cite{heinemann06}. To compute $\Div\mathbf{q}_\mathrm{rad}$, we solve the equation of radiative transport by adopting a grey atmosphere approximation, and by neglecting scattering contributions. 
\EQ
\label{eq:rad}
\hat{{\mathbf{n}}}{\bm \cdot}{\bm \nabla} I\, = \,-\,\kappa_\mathrm{tot}\rho{(I-S)}\,,
\EN
where, {$ \mathbf{I(x,z,t,\hat{n})}$} is the specific intensity along direction $\mathbf {\hat{n}}$. The source function, $S=(\sigma_\mathrm{SB}/\pi)T^4$ is the frequency integrated Planck's function with $\sigma_\mathrm{SB}$ being the Stefan-Boltzman constant. The integration of Eq.~(\ref{eq:rad}) over solid angle, $\varOmega$, will give us $\Div\mathbf{q}_\mathrm{rad}$ by
\EQ
\Div\mathbf{q}_\mathrm{rad}\,=\,\kappa_\mathrm{tot}\rho\oint_{4\pi}(I-S)\,{\mathrm{d}}\varOmega\, .
\EN

For this angular integration we use eight rays, four along $x$ and $z$ axes and along face diagonals of the $x-z$ grid, and correctly scale the angular weight factors for two dimensions \citep{barekat2014near}. We do not use tabulated opacities, rather apply analytical power-law fits to the Rosseland mean opacity functions.  We use solar abundances X=0.7381, Y=0.2485, metallicity Z=0.0134. The bound-free, free-free and $\mathrm{H}^{-}$ opacities are combined to give the total opacity, $\kappa_\mathrm{tot}$. For the optically thin plasma above, we use a cooling function, $Q_\mathrm{thin}=y_\mathrm{H}n_\mathrm{H}^2\Lambda(T)$, where $y_\mathrm{H}$ is the ionization fraction of Hydrogen, and $n_\mathrm{H}$ is the number density of the element Hydrogen. $\Lambda(T)$ is a function of temperature in units of J m$^3$s$^{-1}$ and available in tabulated form, computed from atomic data \cite{Cook89}. The radiation that escapes outward contributes to the cooling of the solar plasma. At the same time, an equal part of the optically thin radiative flux is emitted radially inward and is absorbed in the chromosphere, where it contributes to radiative heating instead of cooling \citep{Carlsson12}. To calculate this heating at every time-step in a 3-dimensional simulation is non-trivial, therefore we approximate  $Q_\mathrm{thin}\rightarrow 0$ smoothly for $T<30,000$\,K. Further, in absence of any self-sustained heating process, the temperature at the top boundary at $z=32$\,Mm is maintained at $10^6$\,K to prevent the atmosphere from collapsing while the solar convection is building up. The saturation of convection takes about 1\,hr of solar time. The results utilized in this work correspond to 25\,min of data taken after the saturation of convection. 

\begin{figure}  
\includegraphics[width=0.8\textwidth]{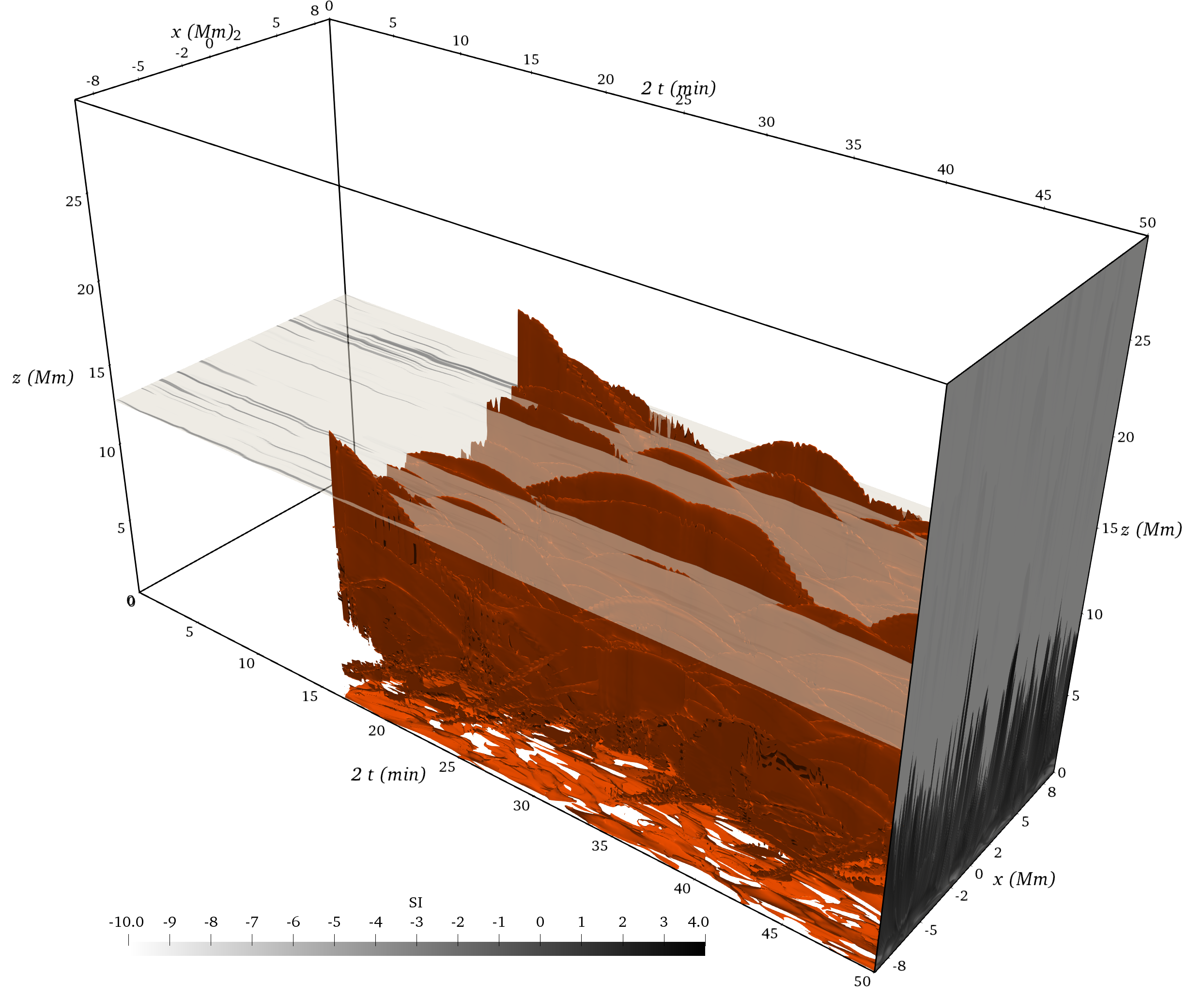}   
\caption{Spicules denoted by the iso-contour (red) of 15,000\,K synthetic emission (SI) in the simulation domain with a resolution of 16\,km for 25\,min of solar time. Note that the time axis is stretched by a factor of 2. The tips of the spicule iso-contours clearly trace parabolic curves in the space-time cube. The horizontal plane at $z=13$\,Mm, shaded by synthetic intensity contours for emission at 15,000\,K, denotes the $x-t$ cross-section where each wiggle is a spicule during the time the spicule cross-section intersects the plane. The vertical $x-z$ plane shows a snapshot of the forest of spicules at $t=25$\,min after the relaxation of the simulation. The imposed vertical magnetic field $B_{imp}$ is 100G.}
\label{fig:paraview}
\end{figure}

\begin{figure} 
\centerline{\includegraphics[width=0.5\textwidth]{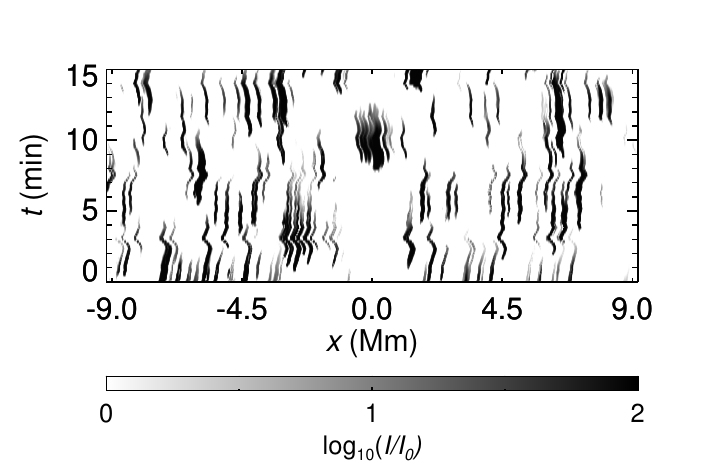}}

        \caption{The horizontal plane at $z=10$\,Mm for the 100\,G simulation setup as in Fig.~\ref{fig:paraview}, where the colorbar is for the synthetic intensity. The gray shaded wiggle-like structures depict the cross-section of the spicules. Note that from here, it can be seen that no spicule lies fully in a single vertical slice, i.e., a $z-t$ slice.}
\label{fig:xt_horiz}
\end{figure}

\subsection{Spicule counting}
\label{S-Num-count}

In order to count the synthetic spicules as seen in emission, we first carry out forward modeling on the 2D simulation data in the $[x,z]$ plane using a simple Gaussian expression. The synthetic intensity for an upper chromospheric or transition region line may be given by,
\begin{equation}
\label{eq:syn}
I_\lambda=\int G_\lambda(\rho, T) \varphi(T) dT.
\end{equation}
This expression is valid for temperatures between $10^4$--$10^8$\,K \citep[see][]{Landi08}, where $$G_\lambda(\rho, T)=\exp\left[-\left(\log(T/T_\lambda)/w\right)^2\right]$$ is the ``contribution'' function for the line with $T_\lambda=15000$\,K and $w=\log1.8$ and $\varphi(T)=(\rho/\overline{\rho})^2 ds/dT$ is the differential emission measure scaled with the square of the horizontally averaged density, $\overline{\rho}(z)$.  Similar expressions have been used by \cite{Iijima17} for depicting synthetic coronal and chromospheric jets, respectively. Normally, the integration is performed along the line-of-sight (LOS), the infinitesimal element along which is $ds$. However, for two-dimensional models, the integration is irrelevant as there is only one grid point along the LOS. Counting of solar spicules is carried out by creating a time-distance data cube from the time evolution of the 2D simulation in $[x,z]$-plane. At a given height, $z$, we take a horizontal cross-section (see Fig.~\ref{fig:paraview}) i.e., parallel to the $[x,t]$ plane and plot the emission contours above a predefined threshold. This procedure yields 2D shaded contour plots featuring distinct wiggle-like structures, which correspond to $x-t$ cross-sections of the simulated spicules as shown in Fig.~\ref{fig:xt_horiz}. At any height $z$, we can then just count the number of distinct wiggle-like features by prompting a threshold brightness of pixels that represent the number of spicules whose height $\ge z$. Essentially, we construct a cumulative distribution function versus a height plot. We take note of this count every equidistant interval of the height. This methodology is repeated for all the simulations with varying imposed vertical magnetic field strengths, i.e., $B_\mathrm{imp}=$10\,G, 50\,G, 100, 200\,G.
\subsection{Parabola fits}
\label{S-Parabola}

After the spicule counting exercise, we shift our attention to kinematics of spicules by studying their trajectories in $[z-t]$ plane that closely resemble parabolas. These trajectories can be used to estimate the deceleration of the spicule tips. In order to obtain a best-fit parabola for a given spicule trajectory in height-time, we once again use the 3D data cubes in $x-z-t$ space.  Because of the kink mode or swaying of the spicule, the trajectory of the tip does not lie in a single $[z,t]$ plane with fixed $x$, but occupies neighboring planes too. Therefore, we visually identify the $x$ and $z$ coordinates of interest; Subsequently, we generate $[z,t]$ cross-sections at these specified $x$ coordinates. To obtain the final images, we employ the Maximum Intensity Projection (MIP) technique \citep{sakas1995optimized}. 

The MIP technique is widely used for deep learning studies for Magnetic Resonance Angiogram (MRA) images \citep{sun1999performance} - a type of MRI applied explicitly to blood vessels. Here, MIP operates by taking a central slice as input and considering 2 or 3 adjacent slices on both sides of the chosen $x$ coordinate. Then, it calculates a resultant slice by projecting the pixel with the highest brightness value from each of the input $[z,t]$ planes onto it. This process ultimately yields a 2D projection of a 3D object. The rationale behind adopting the MIP technique lies in the oscillatory mode of spicules, analogous to human blood vessels. Due to their swaying motion, several spicules span across multiple adjacent $z-t$ planes, rather than being confined to a single slice with given $x$. Consequently, constructing a vertical cross-section at a specific index may not accurately capture the complete shape of the target spicule. The utilization of MIP enhances the comprehensive representation of intricate 3D structures. However, the implementation technique employed by MIP, which generates an overlapped image of cumulative slices, presents challenges in studying shorter spicules. The heightened density of short spicules complicates their analysis. Consequently, to facilitate a more manageable investigation, we have chosen to narrow our focus to spicules surpassing a predefined minimum height criterion. Specifically, our analysis concentrates on spicules lying above the 75$^{th}$ percentile of maximum height. This approach is supported by \cite{suematsu1995high}, who argued that studying longer spicules yields more reliable results, as shorter ones can frequently be mistaken for fibrils and other prevalent structures.

Subsequently, we employ a second-order polynomial, representing a perfect parabolic equation, for fitting purposes. The selection of relatively longer spicules for this fitting is motivated by the observation that the parabolic nature is obscured primarily due to the overlap of spicules at lower heights. The fitted parabola adopts a general equation as 
$$z=-at^2+bt+c,$$ 
where $a,b\geq 0$ will correspond to the acceleration and initial velocity, respectively. For the final step, we visually select seven data points defining the parabola, which the fitting routine uses to determine the best-fit curve. It is worth noting that both visual error and fitting-related error decrease as more data points are incorporated into the fitting process, as expected. However, for practical efficiency, we limit our analysis to a 7-point fit. While efforts have been made to minimize errors, it is crucial to acknowledge that errors persist due to visual inaccuracy, and these errors become more pronounced while fitting shorter spicules that often overlap with each other. \footnote{
Python codes for the Spicule analysis are available at \href{https://github.com/Kartav33/Spicules_Analysis_Codes}{github.com/Kartav33/Spicules\_Analysis\_Codes}} 

\section{Results} 
\label{S-Results} 

\begin{figure}    
\centerline{\hspace*{0.015\textwidth}
         \includegraphics[width=0.5\textwidth,clip=]{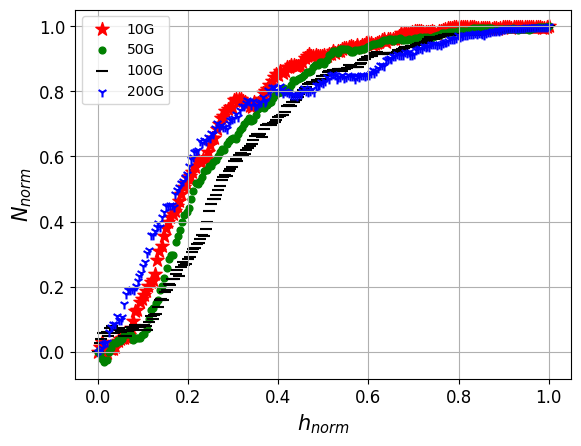}
         \includegraphics[width=0.5\textwidth,clip=]{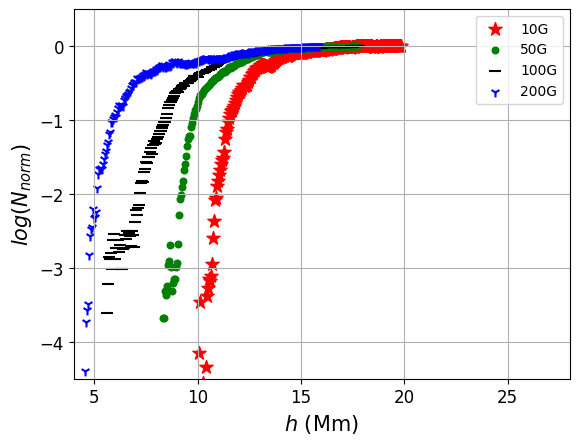}
        }
\vspace{-0.4\textwidth}   
\centerline{\bf      
\hspace{0.0 \textwidth}  \color{black}{(a)}
\hspace{0.47\textwidth}  \color{black}{(b)}
   \hfill}
\vspace{0.37\textwidth}    
\centerline{\hspace*{0.015\textwidth}
         \includegraphics[width=0.5\textwidth,clip=]{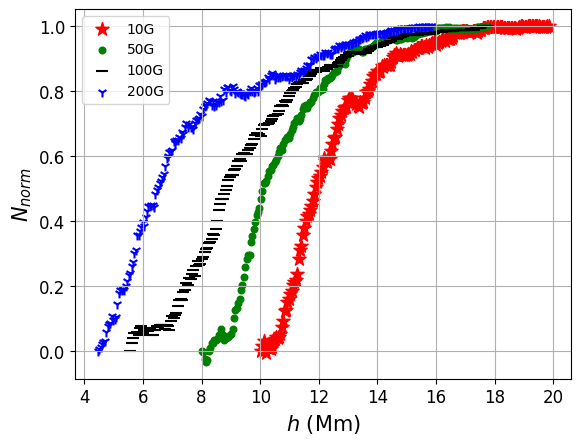}
         \includegraphics[width=0.5\textwidth,clip=]{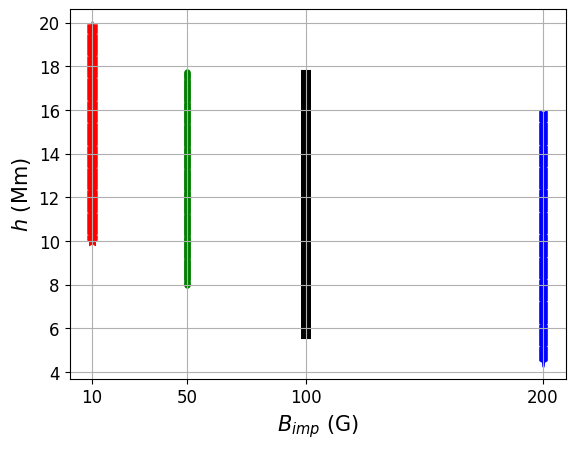}
        }
\vspace{-0.4\textwidth}   
\centerline{\bf     
\hspace{0.0 \textwidth} \color{black}{(c)}
\hspace{0.47\textwidth}  \color{black}{(d)}
   \hfill}
\vspace{0.37\textwidth}    
              
\caption{(a) The normalized cumulative distributive function of spicule count $N_{norm}$ versus normalized height. (b) $N_{norm}$ in logarithmic scale versus height in its original values. (c) Same as the top right panel, key difference being the $N_{norm}$ in normal scale. (d) Height of all the spicules that were studied plotted against the magnetic field imposed for each of the four numerical experiments indicated in the legend.}
\label{fig:F-4panels}
\end{figure}

We use two different methods to visualize our spicule height distribution results. Fig.~\ref{fig:F-4panels}a provides an overview of the overall behavior. This is achieved by normalizing both the height and number of spicules. This normalization process followed is given by Eqs.~(\ref{eq:height})-(\ref{eq:number}). It is worth noting that our method yields the count of spicules longer than a given height where the horizontal cross-sections are taken. However, our definition of the CDF$(h)$ refers to the number of spicules shorter than a given height, $h$. Height is normalised using the formula:
\begin{equation}
\label{eq:height}
    h_\mathrm{norm} = \frac{h- h_\mathrm{min}}{h_\mathrm{max} - h_\mathrm{min}}
\end{equation}
The minimum height has been chosen differently for each imposed vertical magnetic field, $B_\mathrm{imp}$, due to the fact that it is difficult to discern the shortest spicules using our method. Incorporating the cumulative distribution function (CDF) plots, we established minimum heights for analysis, selecting values of 10 Mm, 8 Mm, 5.6 Mm, and 4.45 Mm corresponding to magnetic field strengths of 10G, 50G, 100G, and 200G, respectively. Implying that the minimum height after which a spicule can be distinctly identified, decreases with increasing magnetic field strength as it can be seen in Fig.~\ref{fig:T-spicdata}. These heights signify the point beyond which there is a consistent and monotonic decrease in the spicule count. The non-monotonic behavior observed below these chosen heights may be attributed to challenges in distinctly identifying spicule structures at lower altitudes. The number of spicules longer than a given height, $h$, is normalized using the following:
\begin{equation}
\label{eq:number}
    N_\mathrm{norm}=\frac{N_\mathrm{total}-N(h)}{N_\mathrm{total}},
\end{equation}

where, $N_\mathrm{total}$, is the total number of unique spicules observed during the duration of 30\,min in any simulation. Panel a of Fig.~\ref{fig:F-4panels} illustrates that irrespective of the magnetic field strength present in the chromosphere, the trend of normalized height versus normalized frequency remains largely consistent. This uniformity underscores the common characteristics exhibited by a distribution of solar spicules.

In panels (b) and (c) of Fig.~\ref{fig:F-4panels}, only the CDF on the vertical axis is normalized while keeping the original values of heights in megameters on the horizontal axis. By repeating the above procedure for all the data sets for each $B_\mathrm{imp}$, we obtain the CDF versus height for each case. These plots reveals that the maximum height attained by simulated spicules decreases, and the curve flattens earlier with increasing magnetic field strength. This signifies an inverse relation between the maximum height reached by spicules and the magnetic field strength in the domain.

\begin{figure}
\centering
\includegraphics[width=0.85\textwidth]{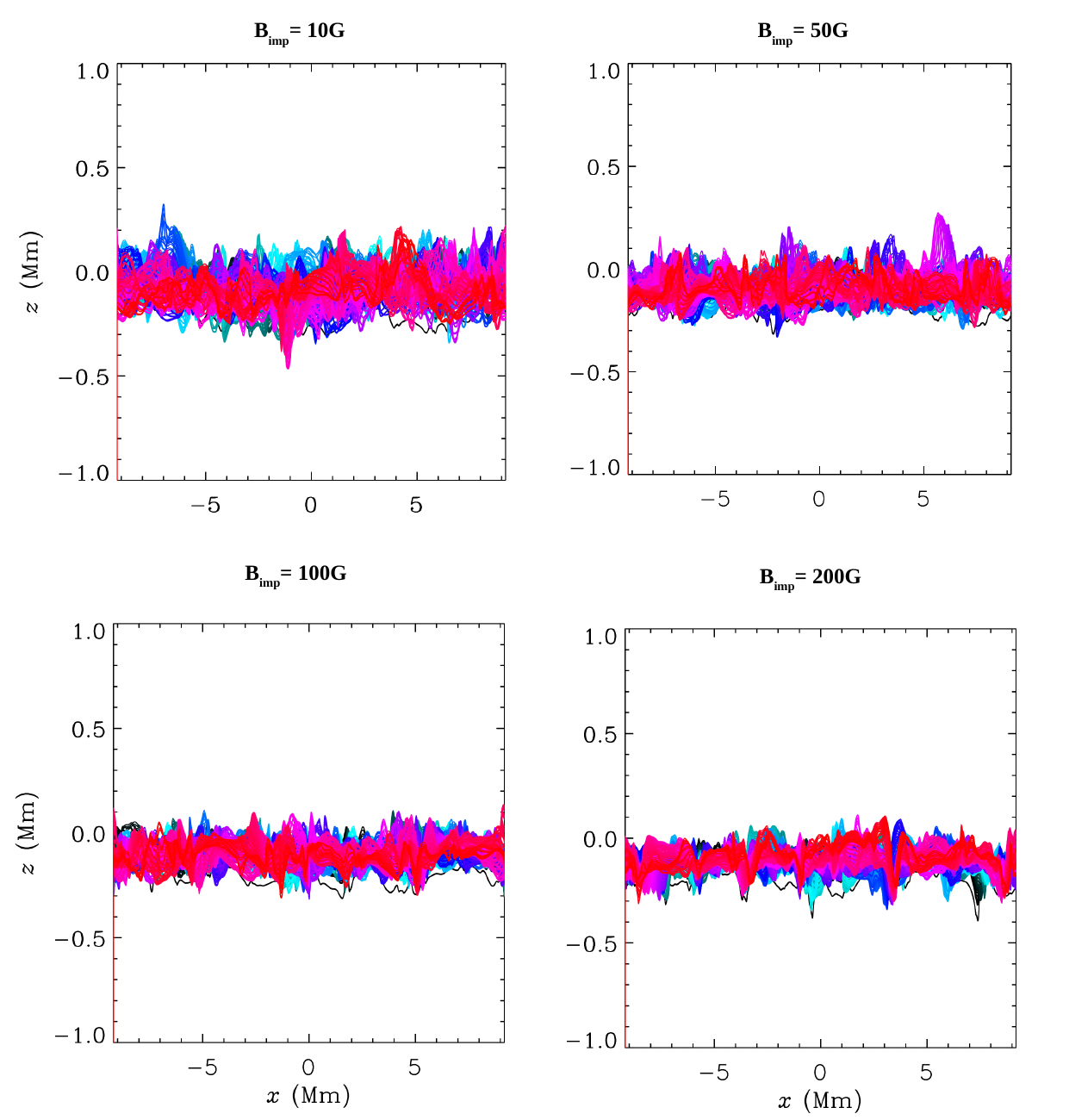}
    \caption{The oscillation of the $\tau=0.1$ surface for data spanning a duration of 25\,min for four different cases of magnetic field. The color of the curve corresponds to its time stamp.}
    \label{fig:tau0.1}
 \end{figure}

Let us explore the possible reasons for this anti-correlation. One possibility is the suppression of the intensity of solar convection in the presence of stronger magnetic fields. The other reason could be the dependence of the coronal pressure on the magnetic field. The higher the coronal pressure, the more challenging will it be for solar spicules to reach higher altitudes in the solar atmosphere. To test the first hypothesis, we look at the warping of the optical depth, $\tau=0.1$ surface over the domain and for 25 min of simulation data. The amplitude of distortion of the $\tau=0.1$ line is shown in Figs.~\ref{fig:tau0.1}a--d for the imposed magnetic field values indicated. 

\begin{figure}
    \centering
\includegraphics[width=0.4\textwidth]{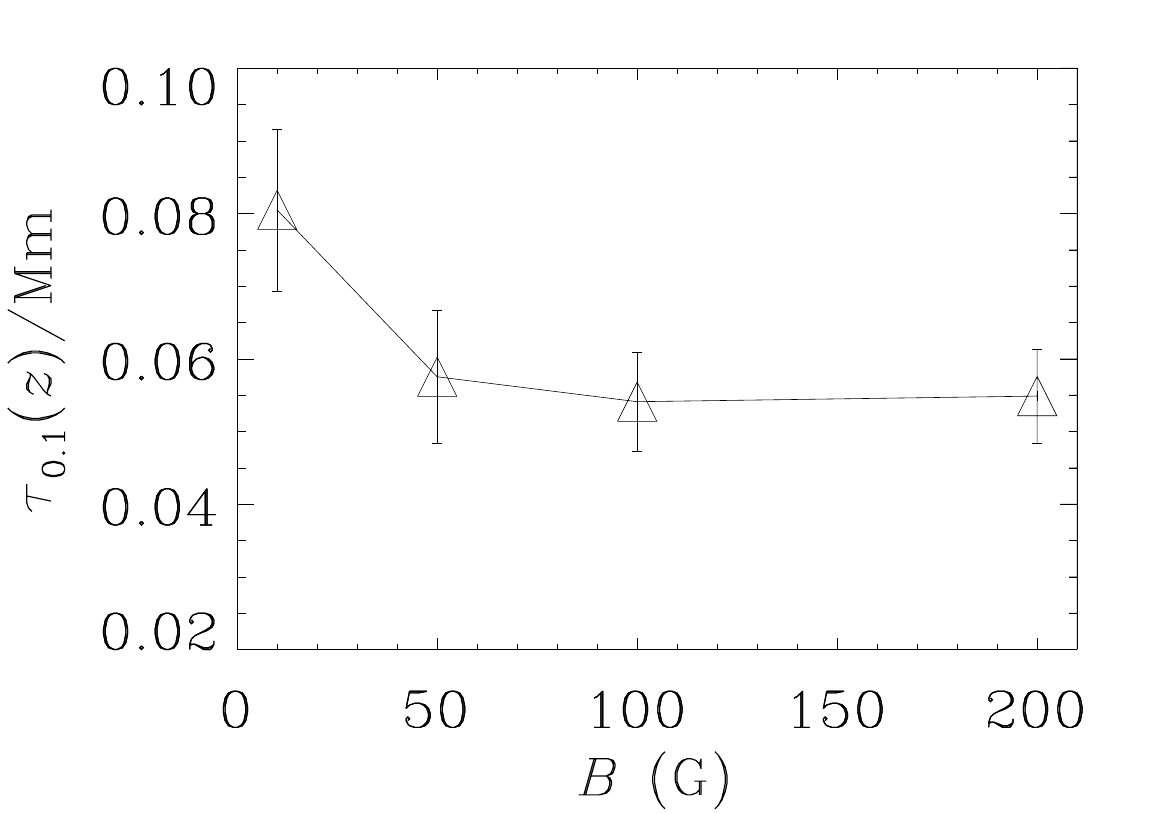}\\
\includegraphics[width=0.4\textwidth]{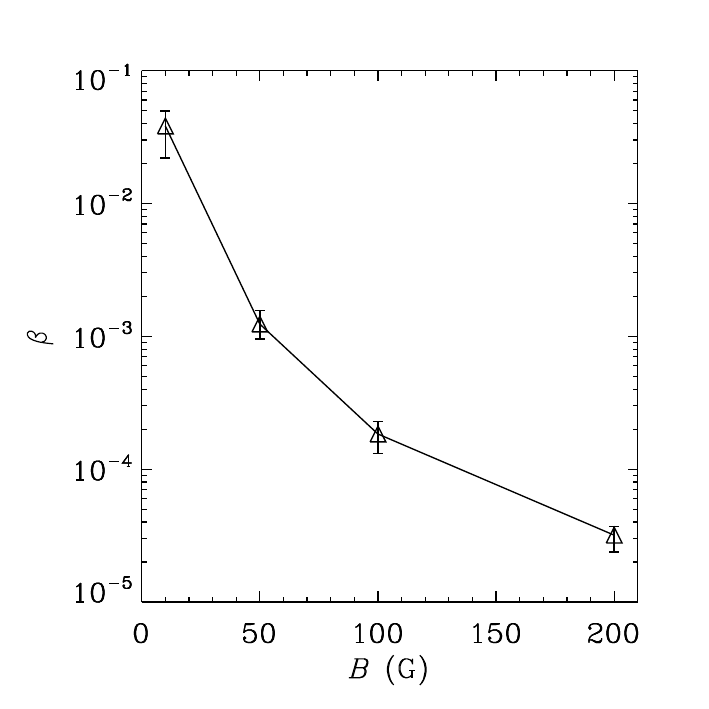}\\
\includegraphics[width=0.4\textwidth]{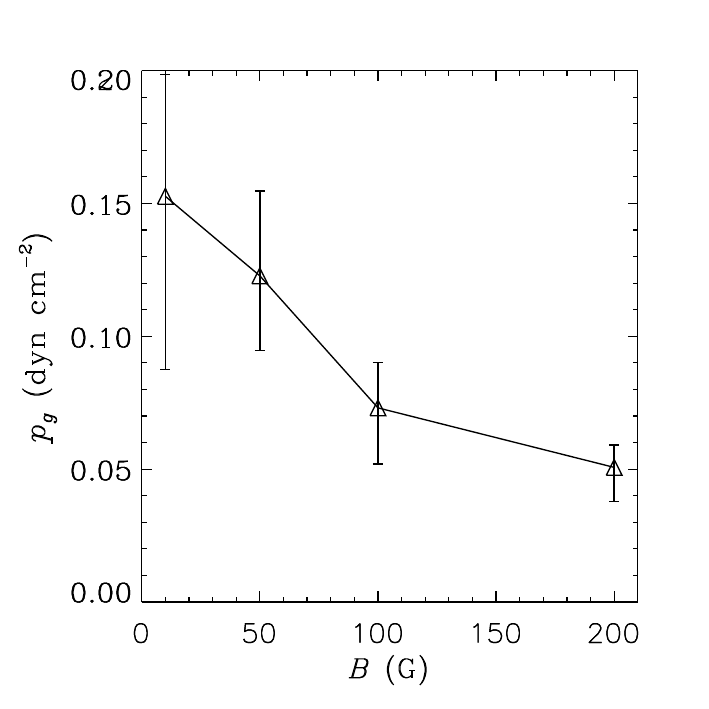}
    \caption{(a) Root-mean-square of the vertical oscillation extent of the $\tau=0.1$ surface averaged over all grid points in the horizontal direction and over 25\,min of data, plotted for the four imposed vertical magnetic field cases shown in Fig.~\ref{fig:tau0.1}. (b) Plasma-$\beta$ and (c) gas pressure in the atmosphere averaged between 8--14\,Mm for different values of $B_\mathrm{imp}$. The error bars denote the spread over 15\,min of data.}
    \label{fig:tau_vs_B}
\end{figure}

We summarize our findings in Fig.~\ref{fig:tau_vs_B}a, where we first note a decrease of the warping amplitude with increasing $B_\mathrm{imp}$ (for lower values of $B_\mathrm{imp}$ i.e., 10--50\,G) and thereafter a levelling. This warping amplitude indicates the amplitude of the quasi-periodic forcing supplied by the convection to propel the lighter chromospheric plasma to form spicules. The decrease in quasi-periodic forcing amplitude leads to shorter spicule heights. This only partially explains the slight decrease of maximum spicule height when the imposed magnetic field increases by a factor of 20. Next, we plot the coronal gas pressure averaged between heights 8--14\,Mm for all four cases at a time when the convection is relaxed and the spicules have formed. 
With {\em increasing} $B_\mathrm{imp}$, we find that, in fact, gas pressure {\em decreases} slightly while the total pressure, $p_g+p_g/\beta$, varies as $B_\mathrm{imp}^2$ as expected for low plasma-$\beta$ regimes (see Figs.~\ref{fig:tau_vs_B}b--c). The decrease in coronal gas pressure should intuitively lead to increased spicule heights due to elevation of the height of the
transition region for a cooler corona \citep{Shibata1982}, as also reported in MHD simulations by \cite{Iijima15}. An elevated transition region implies a lower density of the chromosphere; hence, a given forcing will drive the spicules higher.    
On the contrary, increasing the magnetic field, thereby the Maxwell's stresses, reduces the "breathing" amplitude of the $\tau=0.1$ surface. Thus, in our model, the reduction of coronal pressure and reduction of convection amplitude act in opposite ways on the spicule height. Our results show that the photospheric convective amplitude, as measured by the motion of the $\tau=0.1$ surface, has a dominant effect on the spicule height when $B_\mathrm{imp}$ is varied while keeping the temperature at the top of the boundary fixed at $10^6$\,K.
That may be the reason why upon increasing the magnetic field we see only a slight decrease (rather than a drastic one), in the maximum spicule heights. 
It is also worth mentioning that even though the magnetic pressure is enhanced $400$ times from $B_\mathrm{imp}=10$\,G to 200\,G (Fig.~\ref{fig:tau_vs_B}b), the decrease in spicule heights is not as spectacular. This indicates that in simulations like those used here, the magnetic pressure is not playing a significant role in determining the strength of the slow mode shocks that accelerate the spicule plasma \citep{dey2022polymeric}, and in turn, govern the maximum height attained by the spicule tips.


\begin{figure}    
\centerline{\includegraphics[width=0.4\textwidth]{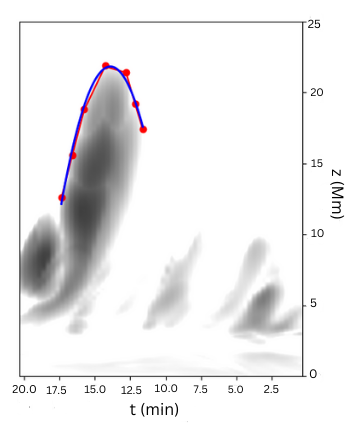}}
\caption{Figure shows the fitted parabola with blue line (smooth line), visually selected points by red dots and the red dotted line, where the equation of the parabola fitted is $z = -0.81t^2 + 22.49t -134.45$, where $t$ is in \,min and $z$ in Mm. This is done for the same test case data from \cite{dey2022polymeric} where $B_{imp}$ is 200\,G. This spicule belongs to the subset that was made under the same criteria as we have used here. {\bf The fading base of the spicule in the synthetic intensity profile is a consequence of the density filtering as described in section~\ref{S-Methodology}}.}
\label{fig:parabola}
\end{figure}. 

\begin{figure}
  \begin{minipage}[c]{0.5\linewidth}
    \centering
    \includegraphics[width=0.75\linewidth]{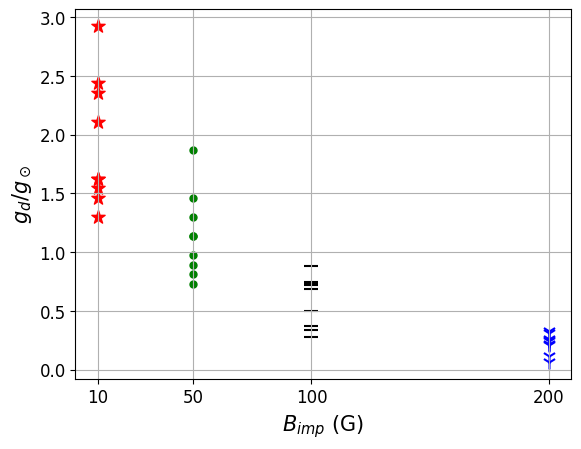}
      \caption{Plot showing the normalized deceleration versus the imposed magnetic field for 8 tallest spicules for each $B_\mathrm{imp}$.}
        \label{fig:F-spicdata}
  \end{minipage}
\end{figure}

Next, we present results from our study of the parabolic nature of spicules tips in $z-t$ space (see Fig.~\ref{fig:parabola}). Here, the parabolic nature is clearly discernible only in the top $\sim25$ percentile of the height distribution. Therefore, a total of 8 spicules were isolated for each of the four cases. For every spicule, we fitted the parabola three times by selecting points manually on the time-distance curve every time and averaged the fitting parameters in order to obtain an average value. In order to compare the deceleration, $g_d$, for the different cases of imposed magnetic fields, we normalized the values by a constant solar gravity of $g_\odot=274$ m\,s$^{-2}$. 
We refer to the Table.~\ref{fig:T-spicdata} for the first few entries for each case. From Fig.~\ref{fig:F-spicdata}, a monotonic decline can be seen in the values of deceleration of spicules with increasing imposed magnetic field. It is worth mentioning that for the weaker magnetic field strength, the deceleration of the longer spicules is much higher than the local solar gravity. The deceleration becomes comparable to $g_\odot$ for the fields lying between 50-100\,G and finally much lower than $g_\odot$ under strong magnetic field strengths.  

\begin{table}[htbp]
  \begin{minipage}[c]{0.5\linewidth}
    \centering
    \begin{tabular}{cccc}
      \toprule
      $B_{imp}$ & $z$ (Mm) & $g_d$ (m\,s$^{-2}$) & $\left|g_d/g_\odot\right|$ \\
      \midrule
      10G  & 19.6 &  645.22 & 2.35 \\
           & 18.8 &  422.73 & 1.54 \\
      50G  & 17.7 &  311.48 & 1.14 \\
           & 17.3 &  511.93 & 1.87 \\
      100G & 17.7 &  102.34 & 0.37 \\
           & 17.4 &  93.44 & 0.34 \\
      200G & 15.8 &  66.75 & 0.24 \\
           & 15.1 &  17.80 & 0.06 \\
      \bottomrule
    \end{tabular}
     \caption{Deceleration, $g_d$, calculated using the leading coefficient of the fitted parabola for every first and second longest spicules in each case. 
     }
  \label{fig:T-spicdata}
  \end{minipage}%
  \end{table}

\section{Discussion} 
      \label{S-Discussion}   

Our 2D rMHD simulations indicate that the strength of the local magnetic field plays a direct role in determining the maximum height of the solar spicules. We identify one possible reason, namely, the suppression of convection at the photosphere by Maxwell's stresses. Moreover, if convection is suppressed, then the "breathing" of the granules in the upper photosphere also decreases, and so does the quasi-periodic kicks to the lighter plasma above in order to form spicular structures. 
Likewise, a higher ambient coronal pressure may also suppress the growth in the height of the spicules, by affecting the strength of the propagating slow mode MHD shocks. Note that in our simulations, the coronal temperature is fixed at $10^6$\,K and does not vary in tandem with the imposed magnetic field. 
The atmosphere above the active region is known to be much hotter than in Quiet Sun regions, which in turn is hotter than Coronal Hole regions. However, in 2D models like those used here, there is no self consistent coronal heating. It remains to be verified if artificially increasing the temperature at the coronal boundary in proportion to the $B_\mathrm{imp}$ causes a further decrease in the spicule heights in our numerical experiments than those depicted in 
Fig.~\ref{fig:F-4panels}d. On the other hand, explaining our result regarding the decrease of the deceleration of the spicules with increasing magnetic field is both non-trivial and counter-intuitive and may have to do with how magnetic tension affects the parabolic shape of the time-distance curve of the spicule tips. 

The MIP technique implemented in this work produces an overlapped image of trajectories of neighboring spicules as well, as previously mentioned, due to which a restricted analysis was carried out by only considering spicules longer than 75 percentile in height distribution. In the future, one can make an improved and modified extension of MIP to study even the shorter spicules that flags off all the neighboring overlapping spicule trajectories that are not of any interest and thus allow us to study the shorter spicules as well. Furthermore, an improvement can be made while analysing the deceleration of the simulated spicules. Instead of fitting a two-dimensional parabola to a (2+1)-dimensional spicule of Fig.~\ref{fig:paraview} that has been collapsed on a single plane using MIP, one can go over fitting a series of 2-dimensional parabola to the actual (2+1)D spicules in the domain, ultimately fitting a curved surface and allowing a topographic study. This will not only provide the deceleration data with much higher accuracy but will also open the possibility of doing a study involving the spicule velocity and its non-parabolic nature. 

The next step should be to compare our theoretical findings with observed spicules near the solar limb by classifying them according to the strength of the underlying photospheric magnetic field strength. Since the surface magnetic field evolves at a much slower rate than spicule lifetimes, in order to reduce projection effects even the magnetic field measured before the active region has reached near the limb may be used. A systematic study can be carried out by combining several near simultaneous data sets from both the Broadband Filter Imager (BFI) for chromospheric images and Narrowband Filter Imager (NFI) for photospheric magnetograms onboard the Hinode spacecraft. 


This work used the DiRAC Data Intensive service (DIaL2) at the University of Leicester (project id: dp261), managed by the University of Leicester Research Computing Service on behalf of the STFC DiRAC HPC Facility (www.dirac.ac.uk). The DiRAC service at Leicester was funded by BEIS, UKRI and STFC capital funding and STFC operations grants. DiRAC is part of the UKRI Digital Research Infrastructure. P.C. acknowledges the allocation of computing
resources at the PDC Center for High Performance Computing at KTH in Stockholm, funded by the National Academic Infrastructure for Supercomputing in Sweden. Computing time provided by Nova HPC at IIA as well as Param Yukti facility at JNCASR under National Supercomputing Mission, Govt. of India is gratefully acknowledged. K.K. was hosted at IIA under their Visiting students program (VSP), when this work was initiated. P.C. acknowledges support from Indo-US Science and Technology Forum (IUSSTF/JC-113/2019). R.E. acknowledges the Science and Technology Facilities Council (STFC UK, grant No. ST/M000826/1), NKFIH (OTKA Hungary, grant No. K142987) and PIFI (China, grant number No. 2024PVA0043) for enabling contribution to this research. This work was also supported by the NKFIH Excellence Grant TKP2021-NKTA-64.
\section{Code Availability}
The MHD setup, along with the parent Pencil Code is publicly available on Zenodo via the following link: \url{https://doi.org/10.5281/zenodo.11481216}. All Python scripts for the Spicule analysis are available at \href{https://github.com/Kartav33/Spicules_Analysis_Codes}{github.com/Kartav33/Spicules\_Analysis\_Codes} 
\bibliography{bibliography}{}
\bibliographystyle{aasjournal}

\end{document}